\documentclass[fleqn,usenatbib]{mnras}

\usepackage[T1]{fontenc}
\usepackage{ae,aecompl}

\usepackage{graphicx}	% Including figure files
\usepackage{amsmath}	% Advanced maths commands
\usepackage{amssymb}	% Extra maths symbols
\usepackage[normalem]{ulem} % editing
\newcommand{\D}{\mathrm{d}}
\newcommand{\Ms}{{\ensuremath{\mathrm{M}_{\sun}}}}
\title[Pristine survivors]{Observational constraints on the survival of pristine stars}

\author[M. Magg et al.]{Mattis Magg$^{1,2}$\thanks{E-mail: mattis.magg@stud.uni-heidelberg.de}, Ralf S. Klessen$^{1,3}$, Simon C. O. Glover$^{1}$, Haining Li$^{4}$\\
$^{1}$Universit\"at Heidelberg, Zentrum f\"ur Astronomie, Institut f\"ur Theoretische Astrophysik, D-69120 Heidelberg, Germany\\
$^{2}$International Max Planck Research School for Astronomy and Cosmic Physics at the University of Heidelberg (IMPRS-HD)\\
$^{3}$Universit\"{a}t Heidelberg, Interdiszipli\"{a}res Zentrum f\"{u}r Wissenschaftliches Rechnen, D-69120 Heidelberg, Germany\\
$^{4}$Key Lab of Optical Astronomy, National Astronomical Observatories, Chinese Academy of Sciences,\\\qquad A20 Datun Road, Chaoyang, Beijing 100101, China\\
}

\date{Accepted XXX. Received YYY; in original form ZZZ}

\pubyear{2019}

\begin{document}
\label{firstpage}
\pagerange{\pageref{firstpage}--\pageref{lastpage}}
\maketitle

\begin{abstract}
There is a longstanding discussion about whether low mass stars can form from pristine gas in the early Universe. A particular point of interest is whether we can find surviving pristine stars from the first generation in our local neighbourhood. We present here a simple analytical estimate that puts tighter constraints on the existence of such stars. In the conventional picture, should these stars have formed in significant numbers and have preserved their pristine chemical composition until today, we should have found them already. With the presented method most current predictions for survivor counts larger than zero can be ruled out.
\end{abstract}


%%%%%%%%%%%%%%%%%%%%%%%%%%%%%%%%%%%%%%%%%%%%%%%%%%

%%%%%%%%%%%%%%%%% BODY OF PAPER %%%%%%%%%%%%%%%%%%

\section{INTRODUCTION}
The first stars, Population III (Pop~III) stars, are thought to form in cosmological minihaloes at redshifts of $z\approx20$, a few 100\,Myr after the Big Bang \citep[see e.g.][]{GloverReview, DayalReview}. Forming in the absence of metals, and therefore from gas with much higher temperature than present day giant molecular clouds, these stars were initially predicted to be very massive \citep[e.g.][]{Bromm99, Abel2000}. Later simulations revealed the fragmentation of protostellar disks and a formation channel for low-mass metal-free stars \citep{Clark11, Greif11b, Stacy14}. However, there is so far no consensus on the initial mass function (IMF) of pristine stars and in particular whether these fragments form and survive. Some simulations, such as those cited above, predict that the formation and survival of low-mass fragments should be relatively common. However, other simulations find that only massive Pop III protostars form and do not predict low-mass fragments to survive in significant numbers \citep{Hosokawa11, Hirano14}.

As these stars are born at high redshifts, it is not possible to directly observe the formation of these stars. Furthermore, low-mass Pop~III stars exhibit only very weak feedback. As they do not explode as supernovae (SNe) they will not leave an imprint in the abundance patterns of second generation stars and are therefore invisible to the conventional stellar archaeology approach \citep{Ishigaki18} or searches for the first SNe \citep{Hummel12, Hartwig18b}. These stars are also not predicted to be very luminous, making 21-cm tomography insensitive to their existence \citep{Schauer19}. Thus, the remaining hope for constraining the existence and abundance of low-mass Pop~III stars is the prospect of observing them directly in the local Universe.

Many observational studies have targeted the most metal-poor stars to be found in our Galaxy, with the goal of understanding its early enrichment history \citep{Beers2005, Frebel15, Ishigaki18}. Among these stars we focus on the categories of extremely metal-poor (EMP) and ultra metal-poor (UMP) stars. We refer to stars with metallicities\footnote{For denoting metallicities, we use the standard notation \\$\mbox{[Fe/H]} = \log_{10}(m_\mathrm{Fe}/m_\mathrm{H})-\log_{10} (m_{\mathrm{Fe},\sun} /m_{\mathrm{H},\sun})$\\ where $m_{\mathrm{Fe}}$ and $m_{\mathrm{H}}$ are the mass abundances of iron and hydrogen, and $m_{\mathrm{Fe},\sun}$ and $m_{\mathrm{H},\sun}$ are the respective solar abundances.} of [Fe/H]<-3 as EMP stars and to stars with a metallicity of [Fe/H]<-4 as UMP stars. Note that in this definition UMP stars are always EMP stars too. 

\section{COMPUTING DETECTION PROBABLITIES}
\subsection{Basic idea}
The idea to constrain the existence of surviving pristine stars by statistically analysing non-detections is not new. There are two common methods: \citet{Hartwig15b} and \citet{Ishiyama16} have estimated the probability to find Pop~III survivors in a blind survey and have stated required sample sizes to falsify the predicted numbers at various confidence levels. However, this approach only looks at blind surveys and does not take into account that observers are actually more efficient in finding metal-poor stars than a blind survey. \citet{Oey03}, \citet{Tumlinson06} and \citet{Salvadori07} have considered implications of non-detections based on detected numbers of stars, but do not discuss the statistical significance of their statements.

Our goal is to understand the implications of the non-detection of metal-free stars by the community until today. A key problem in this is the large non-uniformity in selection criteria employed by the observations. Our approach is based on the following idea: the formation sites of extremely and ultra metal-poor stars are thought to be similar to those of metal-free stars. They both are expected to form in high-redshift mini- and atomic cooling haloes. Therefore, we are not aware of any reason why their spatial distribution, ages, luminosity or magnitudes in broad band filters should be significantly different. This idea is confirmed by the simulations of \citet{Starkenburg17a}, where metal-poor and metal-free stars show very similar spatial distributions. Consequently, we assume that on average extremely metal-poor stars are equally likely to be detected as Pop~III survivors. The detection of each EMP star known to the community can be seen as randomly drawing a star from a sample that contains all EMP stars as well as the hypothetical metal-free stars. Therefore, the probability of metal-free stars randomly escaping detection thus far can be computed as
\begin{equation}
\label{eq:prob}
 P_\mathrm{0} = \left(1-\frac{N_\mathrm{surv}}{N_\mathrm{EMP, tot}}\right)^{N_\mathrm{EMP, obs}},
\end{equation}
where $N_\mathrm{EMP, tot}$ is the total number of EMP stars in the galactic halo, $N_\mathrm{EMP, obs}$ is the number of such stars that are observed and $N_\mathrm{surv}$ is the total number of surviving metal-free stars in the halo. We note that $N_\mathrm{EMP, tot}$ includes all stars with a metallicity below $\mbox{[Fe/H]}=-3$ and therefore in particular UMP stars and potential metal-free stars. A similar estimate is made for UMP stars.

As stated before, \citet{Oey03}, \citet{Tumlinson06} and \citet{Salvadori07} have performed similar investigations and concluded that low-mass Pop~III stars must form rarely and that the pristine IMF must be very different from the one observed in the present day Universe. \citet{Tumlinson06} introduced the parameter $F_0$ which is the fraction of Pop~III stars among all stars with a metallicty below [Fe/H]<-2.5. They compute the upper limit from the total number of observed stars in this metallicity range $N_\mathrm{-2.5, obs}$ to be
\begin{equation}
 F_0 < \frac{1}{N_\mathrm{-2.5, obs}}.
\end{equation}
This method does not allow one to draw any conclusions about the statistical significance of their approach. For example, if we assume a detection rate that is twice the ``upper limit'' derived with this method and use that number in Equation \eqref{eq:prob}, we find for $N_\mathrm{-2.5, obs}$ a probability of non-detection of
\begin{equation}
 P_\mathrm{0} = \left(1-\frac{2}{N_\mathrm{-2.5, obs}}\right)^{N_\mathrm{-2.5, obs}} \approx \frac{1}{e^2} = 13.5\,\%.
\end{equation}
Thus the ``upper limit'' of $F_0$ could be violated by a large margin while still being consistent with the observations it has been derived from. \citet{Oey03} and \citet{Salvadori07} use a similar method.

Our approach goes beyond simply looking at the fraction of stars and instead considers the full probability distribution, thus allowing us to determine the full significance intervall. It also enables us to easily include different metallicity ranges, as we will demonstrate below.

\subsection{Number of Pop~III survivors}
\begin{figure}
 \includegraphics[width=0.95\linewidth]{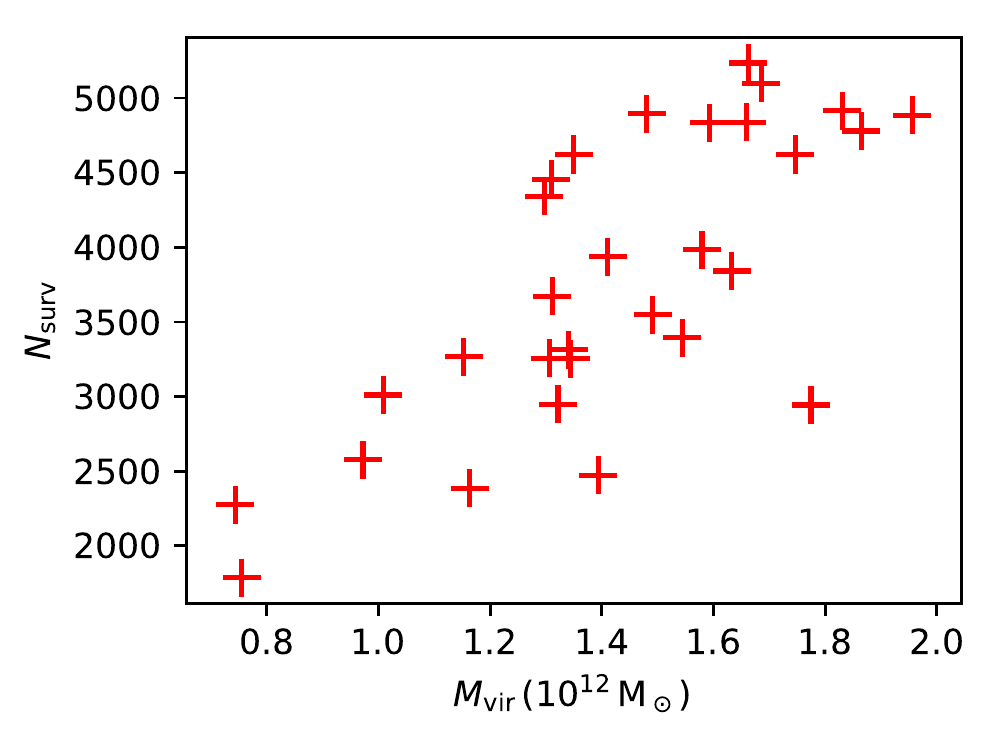}
 \caption{\label{fig:surv}Number of surviving pristine stars in the MW for the 30 MW-like haloes as function of their virial mass. The numbers of these stars were computed by \citet{Magg18} on the basis of merger trees from \citet{caterpillar}.}
\end{figure}

We use 30 different numbers of pristine survivors in the Milky Way (MW) presented in Fig. \ref{fig:surv}. \citet{Magg18} computed these numbers with a semi-analytical model, which simulates early star formation in merger trees of MW-like haloes. The merger trees used for this purpose are from the high-resolution cosmological N-body simulations from the Caterpillar project \citep{caterpillar}. The assumed IMF in this model is a logarithmically flat IMF in the mass range 0.6\--150\,\Ms. This IMF is very top-heavy and it gives many fewer surviving stars than, for example, a Salpeter-like IMF in the same mass range. The number-count of surviving pristine stars spans a range between 1800 and 5200 stars. It is extremely difficult to estimate the error for the survivor count in each individual of the 30 modelled haloes. We assume that the main source of errors is encompassed in the different merger histories of the 30 haloes, i.e., that the 30 survivor counts represent the probability distribution of expected survivor counts in the MW. Even the largest number of surviving pristine stars we adopt is by a factor of two smaller than the one found in \citet[][10000 survivors]{Hartwig15b} on which the \citet{Magg18} study is built. This difference is primarily caused by the significantly improved modelling of feedback and in particular by taking into account the positions of the haloes and the effects of ionizing radiation. The adopted numbers of Pop~III survivors are also by more than order of magnitude smaller than the one in \citet[][around 100000 survivors]{Ishiyama16}. This discrepancy is mostly caused by assuming around an order of magnitude more survivors per Pop~III forming minihalo. \citet{Komiya16} predicted around 3000 Pop~III survivors in the MW, similarly to our estimate. Therefore, our estimate of the number of surviving pristine stars is among the more conservative predictions that still allow for surviving Pop~III stars. \citet{Salvadori07} and \citet{deBennassuti17} also investigate the possiblity of surviving Pop~III stars in the MW and make predictions for how common they are. As they only derive the fraction of a certain group of stars that should be metal-free and do not give absolute numbers, direct comparisons to the survivor counts are difficult. We will investigate the consistency of their predictions with the current state of observations below.

\subsection{Total number of EMP stars in the halo}
To compute with which probability a number of metal-free stars could have escaped detection until now, we have to determine the size of the ``haystack'', i.e., the total number of EMP/UMP stars among which we have to look for pristine survivors. In order to do this we assume that all EMP, UMP and hypothetical pristine stars are located in the stellar halo of the MW. Simulations and semi-analytical modelling confirm that most of surviving metal-free stars should be found in the stellar halo \citep{Hartwig15b, Starkenburg17a}. \citet{Sestito2019} recently found that the majority of known UMP stars indeed are observed in the stellar halo. Therefore, our estimates of the total amount should be appropriate. Furthermore this assumption only enters in our estimate of the size of the ``haystack'', i.e., our model is independent of it as long as it does not lead to us significantly underestimating the total number of EMP/UMP stars in the MW.

We assume a stellar halo mass of $M_{*,\mathrm{halo}} = 10^9\,\Ms$ \citep{Bell2008}. We convert this to a number of stars by assuming an average stellar mass of a typical old metal-poor main sequence turn-off star, i.e. $M_* = 0.6\,\Ms$. This is again a conservative estimate, which probably significantly overestimates the number of EMP and UMP stars. Thus it gives a relatively large ``haystack'', assuming that all stars are at the smallest mass where they can still be detected and the metallicities constrained reasonably well. For the same reason the lower limit of the pristine IMF was selected to be 0.6\,\Ms\ in \citet{Magg18}. Recently, \citet{Youakim2017} estimated that 1/800 of halo stars are EMP stars and 1/80000 of halo stars are UMP stars. Thus, we estimate the total number of EMP and UMP stars to be
\begin{equation}
 N_\mathrm{EMP, tot} = \frac{M_{*,\mathrm{halo}}}{800 M_*} = 2.08\times 10^6
\end{equation}
and
\begin{equation}
 N_\mathrm{UMP, tot} = \frac{M_{*,\mathrm{halo}}}{80000 M_*} = 20800.
\end{equation}
While these numbers are subject to large uncertainties they are the best estimates available to us. To be conservative, we also use a very pessimistic conversion factor between stellar mass and the number count of stars. One source of uncertainty is that \citet{Youakim2017} derived those numbers for stars in the magnitude range $14 < V < 18$ and that EMP and UMP stars could be more common in the more metal-poor outer stellar halo. Therefore, we will discuss below how our results would change, if we underestimated the number of EMP and UMP stars by a factor of two.

\subsection{Number of detected stars}
For the number of observed UMP stars we use the $N_\mathrm{UMP, obs}=42$ stars from \citet{Sestito2019}. We determine the number of detected EMP stars to be $N_\mathrm{EMP, obs}=532$ by a query of the SAGA\footnote{\url{http://sagadatabase.jp}, accessed on 22.01.2019} database \citep{Saga}. Of these stars, 507 are in the metallicity range $-4 <\mbox{[Fe/H]} < -3$. With this method, we are underestimating the number of detected EMP stars as we do not include e.g. the recent detections from TOPoS \citep{TOPos5} or LAMOST (\citealt{LAMOSTDR3}, Li et al. in prep.).

\section{RESULTS AND DISCUSSION}
\begin{figure}
 \includegraphics[width=0.95\linewidth]{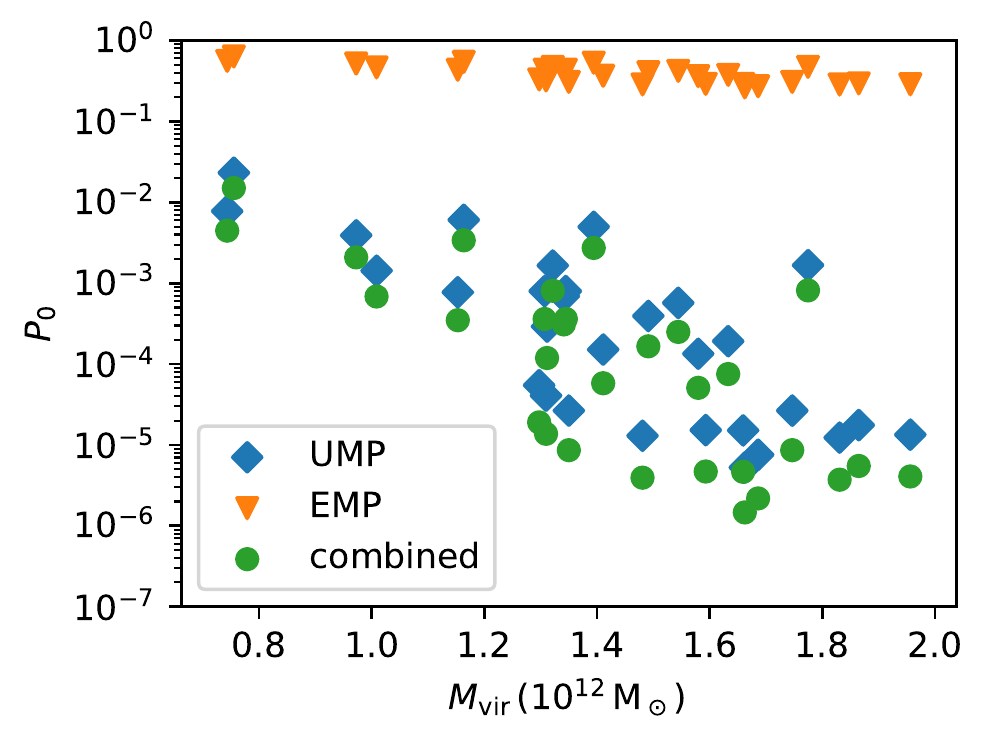}
 \caption{\label{fig:P0} Probability of not detecting metal-free stars until today, as function of halo mass. We show the probabilities derived from UMP (blue diamonds) and EMP (orange triangles) star detections separately and combined (green circles). The UMP stars give much tighter constraints than the EMP stars.}
\end{figure}

By using the estimated numbers of survivors, EMP and UMP stars and their detections, we can compute the probability of the non-detection of these metal-free stars until today with Eq. \eqref{eq:prob}. The probabilities for individual haloes are presented in Fig. \ref{fig:P0}. The non-detection probabilities derived from EMP stars are typically around 50 per cent, while the values derived from the UMP star sample are orders of magnitude smaller. If we remove the UMP stars from the detected EMP sample, i.e., using $N_\mathrm{EMP, obs}=507$, the two non-detection probabilities become independent. We can combine them by multiplication. These combined probabilities offer a slight improvement over the estimate derived from detections of UMP stars. %We combine the probabilities from EMP and UMP stars by multiplying them. However, for this operation we use $N_\mathrm{EMP, obs}=507$, to exclude UMP stars from the EMP estimate and make the probabilities independent. 
As the 30 simulated merger trees represent our a priori distribution of survivor counts, we compute final non-detection probabilities in the MW by averaging the 30 probabilities. We find
\begin{itemize}
 \item $P_{0,\mathrm{EMP}} = 0.39$ for EMP stars,
 \item $P_{0,\mathrm{UMP}} = 0.0019$ for UMP stars and
 \item $P_{0,\mathrm{comb}} = 0.0011$ combined.
\end{itemize}
Thus the estimate of the number of metal-free survivors can excluded at a 99.9 per cent confidence level. At 98 per cent confidence level, this model allows for no more than 1650 metal-free stars. This also rules out the predicted number of survivors from \citet{Hartwig15b} and \citet{Ishiyama16}. The non-detection probabilities derived from EMP stars are typically around 50 per cent, while the values derived from the UMP star sample are orders of magnitude smaller. Even if we underestimate the number of EMP and UMP stars by a factor of two, the predictions from \citet{Magg18} are still incompatible with our estimate at a confidence level of 97 per cent.

While the study by \citet{deBennassuti17} does not include an absolute number of surviving stars, they state that for one of their IMFs, they find that 0.15 per cent of EMP stars should be metal-free. For this prediction they use a Larson-type IMF with the shape
\begin{equation}
 \frac{\D N}{\D m} \propto m^{-2.35} \exp\left(\frac{-0.35\,\Ms}{m}\right),
\end{equation}
between mass limits of 0.1\,\Ms\ and 300\,\Ms. If we continue to assume that one per cent of EMP stars are UMP stars, we can use our formalism to determine the probability of non-detection to be
\begin{equation}
 P_0 = (1-0.0015)^{507} (1-0.15)^{42} = 0.05\,\% ,
\end{equation}
ruling out their assumed IMF with 99.95 per cent confidence. However, it is not clear whether our assumption of the ratio between UMP and EMP stars is consistent with their metallicity distribution function.

\citet{Salvadori07} investigate three cases of different critical metallicities, at which the transition from the primordial to the present day IMF occurs.  They result in $F_0 = 7.5\times 10^{-3}$ for a critical metallicity of $Z_\mathrm{cr}=0$ and $F_0 \approx 7.5\times 10^{-9}$ for $Z_\mathrm{cr}=10^{-4}\,\mathrm{Z}_\odot$ and $Z_\mathrm{cr}=10^{-6}\,\mathrm{Z}_\odot$, where $\mathrm{Z}_\odot$ is the solar metallicity. As the former value was already ruled out by the observations at the time of publication and the latter two are extremely low values, no further conclusions can be drawn from applying more recent observations to their predictions.

In addition to upper limits, we can compute our best estimate of the number of surviving metal-free stars. At merely $N_\mathrm{surv}=300$ the non-detection of metal-free stars until today becomes equally likely to their detection (i.e. $P_{0,\mathrm{comb}} = 0.5$).

These findings imply either that low mass Pop~III stars must be rarer than suggested by \citet{Magg18} or a significant number of them must accrete metals from their surroundings during their life. We will briefly discuss both possibilities here.

A possible explanation for the non-detection of metal-free stars is that they could have been enriched with accreted metals from the interstellar or intergalactic medium during their lifetime, and are now detected among the metal-poor stars. Whether such accretion can occur at a sufficient level to explain the non-detection of metal-free stars is still under debate. \citet{Frebel09} found that metal-poor stars can only very inefficiently accrete metals, while \citet{Johnson11} find accretion can be efficient, if it is not prevented by stellar winds. If enrichment is efficient, \citet{Komiya16} predicted that around 100 -- 170 stars may escape their formation sites before being enriched and be accreted onto the MW stellar halo at later times. However, such stars may have a different distribution of orbits and can therefore not be investigated with the model presented in this study. More recently \citet{Tanaka17} and \citet{Suzuki2018} pointed out that magnetic winds and hot coronae of Pop~III survivors may prevent accretion of metals from the insterstellar medium altogether. Accretion of compact interstellar comets has been investigated as source of metal pollution by \citet{Tanikawa2018}, and the contribution was found to be negligible in most cases.

A second possible pathway that leads to the pollution of surviving Pop~III stars is mass overflow in binaries. For example, \citet{Suda2004} and \citet{Lau2007} have shown that under the right conditions significant pollution with carbon can occur in Pop~III binaries. Observationally, \citet{Arentsen2019} suggest a connection between carbon enhancement and binarity in EMP stars. Such mass transfer would be sensitive to the binarity and the separation of Pop~III stars, which is so far not well constrained. However, in most cases binary mass transfer is insufficient as source of enrichment, as it cannot explain the presence of iron in these stars. Therefore, in order to explain the lack of observed surviving metal-free stars by pollution, the need for accretion from the interstellar medium remains.

The most obvious explanation for the non-detections of Pop~III survivors, would be that metal-free stars with masses below 0.8\,\Ms\ do not form at all or are much rarer than predicted. This would mean that the pristine IMF must either be more top heavy or truncated towards lower masses. We perform a simplistic estimate here of how much we would need to change the IMF form the one assumed in \citet{Magg18} in order to arrive at survivor numbers consistent with observations. \citet{Magg18} used a logarithmically flat IMF in the mass range 0.6\--150\,\Ms. Such an IMF produces on average one surviving star per 520\,\Ms\ of forming Pop~III stars, where we assumed that these stars survive in the range of 0.6\--0.8\,\Ms. For comparison, the present day IMF from \citet{KroupaIMF} predicts around one star in this mass range per 10\,\Ms\ of stars formed. The assumed pristine IMF leads to the prediction of an average of 3750 surviving Pop~III stars. At fixed star formation efficiency and feedback efficacy, to scale this number down to the above given upper limit of 1650 survivors, we would need to reduce the number of survivors to one per 1200\,\Ms\ of Pop~III stars. This could either be achieved by raising the lower limit of the IMF to 0.7\,\Ms\ or by changing its slope from $\alpha = -1.0$ (i.e., logarithmically flat) to $\alpha = - 0.8$. In particular, these constraints are strongly inconsistent with Pop~III stars forming with an IMF similar to the one observed in the present day. Without pollution of almost every low-mass Pop~III star, an IMF that, compared to the present day IMF, is either very top-heavy or truncated towards sub-solar masses is required.

The largest uncertainty in our method lies in our estimate of the total size of the ``haystack'', i.e. the total number of EMP and UMP stars in the Milky Way and in particular in the outer stellar halo. While the numbers used in this paper are uncertain we chose a conservative estimate. Future surveys that allow to more completely and deeply explore the outer stellar halo will reduce these uncertainties. Additionally, the number of UMP and in particular EMP stars cited in this study are lower limits, as the SAGA catalogue is not entirely up-to-date and because there are more processed, but not yet published detections of EMP stars (Li et al. in prep).

In summary, we derived new upper limits on the number of surviving metal-free stars in the MW. We conclude that such stars must form very rarely or else have been polluted by metals during their lifetime. We demonstrate the ability of our approach to constrain the primordial IMF by non-detections and highlight the need for a lower mass cut-off of the IMF or an even more top-heavy functional form than adopted here, should pollution of these stars not be significant. Future surveys and individual detections of EMP and UMP stars will further strengthen constraints on the survival of metal-free stars.

\section*{Acknowledgements}
We thank the referee for their helpful comments. The authors would like to thank Tilman Hartwig, Anna Schauer, Else Starkenburg and Miho Ishigaki for productive and helpful discussions during the preparation of this letter. We thank Takuma Suda for making the SAGA database publicly available. MM was supported by the Max-Planck-Gesellschaft via the fellowship of the International Max Planck Research School for Astronomy and Cosmic Physics at the University of Heidelberg (IMPRS-HD). HL was supported by NSFC grant No. 11573032. SCOG and RSK acknowledge funding from the Deutsche Forschungsgemeinschaft (DFG) via SFB 881 ``The Milky Way System'' (subprojects A1, B1, B2 and B8). This study is supported by the DFG via the Heidelberg Cluster of Excellence \textit{STRUCTURES} in the framework of Germany's Excellence Strategy (grant EXC-2181/1 - 390900948).


\bibliographystyle{mnras}
\bibliography{lit}

% Don't change these lines
\bsp % typesetting comment
\label{lastpage}
\end{document}